\documentclass[aps,prd,twocolumn,superscriptaddress,nofootinbib,amsmath,amssymb]{revtex4}

\usepackage{hyperref}
\usepackage{amsmath,amssymb,amsfonts}
\usepackage{color}
\usepackage{graphicx}
\usepackage{enumerate} 
\usepackage{colordvi} 
\usepackage{bm}
\usepackage{multirow}

\newcommand{\rockstar}{\texttt{\footnotesize ROCKSTAR} }
\newcommand{\darkquest}{\texttt{D{\footnotesize ARK} Q{\footnotesize UEST} }}
\newcommand{\himpc}{{\hbox {$~h^{-1}$}{\rm ~Mpc}}}
\newcommand{\higpc}{{\hbox {$~h^{-1}$}{\rm ~Gpc}}}

\newcommand{\hiMsun}{{\hbox {$~h^{-1}M_\odot$}}}

\newcommand{\hmpcci}{{\hbox {$~h^3{\rm ~Mpc}^{-3}$}}}

\newcommand{\bfr}{\boldsymbol{r}}
\newcommand{\bfq}{\boldsymbol{q}}
\newcommand{\bfk}{\boldsymbol{k}}
\newcommand{\be}{\begin{equation}}
\newcommand{\ee}{\end{equation}}
\newcommand{\nn}{\nonumber}
\newcommand{\Mh}{M_{\rm h}}
\newcommand{\bMh}{\bar{M}_{\rm h}}
\newcommand{\DA}{D_{\rm A}}
\newcommand{\DAf}{D_{\rm A,fid}}
\newcommand{\Hf}{H_{\rm fid}}

\newcommand{\As}{A_{\rm s}}
\newcommand{\Om}{\Omega_{\rm m}}
\newcommand{\Ok}{\Omega_{\rm k}}
\newcommand{\Ob}{\Omega_{\rm b}}
\newcommand{\ob}{\omega_{\rm b}}
\newcommand{\obs}{{\rm obs}}
\newcommand{\dof}{{\rm dof}}

\title{Cosmological constraints from $N$-body simulations for future spectroscopic galaxy surveys at $2\leq z\leq 3$}

\newcommand{\nthu}{Institute of Astronomy and Department of Physics, National Tsing Hua University, Hsinchu 30013, Taiwan}
\newcommand{\asiaa}{Academia Sinica Institute of Astronomy and Astrophysics (ASIAA), No. 1, Section 4, Roosevelt Road, Taipei 106216, Taiwan}
\newcommand{\ipmu}{Kavli IPMU (WPI), UTIAS, The University of Tokyo, Kashiwa, Chiba 277-8583, Japan}
\newcommand{\kyosan}{Department of Astrophysics and Atmospheric Sciences, Faculty of Science, Kyoto Sangyo University, Kyoto 603-8555, Japan}
\newcommand{\yitp}{Center for Gravitational Physics and Quantum Information, Yukawa Institute for Theoretical Physics, Kyoto University, Kyoto 606-8502, Japan}
\newcommand{\kek}{Theory Center, Institute of Particle and Nuclear Studies,
High Energy Accelerator Research Organization (KEK), Tsukuba, Ibaraki 305-0801, Japan}

\date{\today}

\begin{document}

\title{
Constraining cosmology with $N$-body simulations for future spectroscopic galaxy surveys at $2\leq z\leq 3$
}

\author{Sy-Yun Pu}\email{psyn@gapp.nthu.edu.tw}
\affiliation{\asiaa}
\affiliation{\nthu}

\author{Teppei Okumura}\email{tokumura@asiaa.sinica.edu.tw}
\affiliation{\asiaa}
\affiliation{\ipmu}

\author{Chian-Chou Chen}\email{ccchen@asiaa.sinica.edu.tw}
\affiliation{\asiaa}

\author{Takahiro Nishimichi}
\affiliation{\kyosan}
\affiliation{\yitp}
\affiliation{\ipmu}

\author{Kazuyuki Akitsu}
\affiliation{\kek}

\begin{abstract}
Determining the spatial curvature ($\Ok$) independent of cosmic
microwave background observations plays a key role in revealing the
physics of the early universe. The Hubble tension is one of the most
serious issues in modern cosmology. We investigate halo catalogs
identified from $N$-body simulations at $z=2$ and 3, mimicking
high-redshift galaxy surveys. We measure redshift-space correlation
functions of halos from the two snapshots. We detect clear features of
baryon acoustic oscillations and redshift-space distortions. We find
that we can obtain a few percent constraints on both the geometric
distances and growth of structure at the distant universe in future
surveys.  By taking into account the information of the underlying
matter power spectrum, we demonstrate that we can also achieve
constraint on the Hubble constant $H_0$ with a few percent as well as
the spatial curvature with $|\Ok|\lesssim 0.1$ by observing galaxies
with the number density with $\bar{n}_{\rm g}\simeq 10^{-4}
(\hmpcci)$. Our analysis provides a timely forecast for the upcoming
spectroscopic surveys, which target emission line galaxy or dusty
star-forming galaxy samples.
\end{abstract}

\maketitle



{\it Introduction.}  Cosmological models have been tested via various
observations including cosmic microwave background (CMB)
\cite{Hinshaw:2013,Planck-Collaboration:2016}, gravitational lensing
\cite{DES_Collaboration:2018,Hikage:2019}, galaxy redshift surveys
\cite{Ivanov:2020,dAmico:2020,Kobayashi:2022}, Type-Ia supernovae
\cite{Suzuki:2012,DES_Collaboration:2024}, etc.  Cold dark matter with
a cosmological constant, the $\Lambda$CDM model, has successfully
explained all of these observations so far.  However, there exist
discrepancies between some parameters determined from the early and
local universe, known as the Hubble tension \citep{Verde:2019} and the
$S_8$ tension \cite{Abdalla:2022}. The Dark Energy Spectroscopic
Instrument (DESI)
\cite{DESI_iv:2025,DESI-Collaboration:2025iii,DESI-Collaboration:2025vi},
the latest galaxy survey, reported cosmological constraints that
support dynamical dark energy with $w>-1$. However, combining them
with other datasets still includes the cosmological constant ($w=-1$)
within $2\sigma$ in a large cosmological parameter space
\cite{Choudhury:2024}. Although the geometry of the universe has been
found flat by the CMB, the flatness of the universe has not been
constrained in late universe observations
\citep{Takada:2015,Terasawa:2022,Anselmi:2023,Stevens:2023,Terasawa:2024}. One
way to tackle these issues is to extend the observations of
cosmological objects to the more distant universe \cite{Okumura:2016}.
Cosmological spectroscopic experiments at $z\gtrsim2$ have the
potential to put meaningful constraints on curvature
\citep{Takada:2015}, neutrino mass \citep{Takada:2006,Dvorkin:2019},
inflation \citep{Takada:2006}, and dark energy \citep{Ferraro:2019}.

Indeed, in the coming years a number of optical and near-infrared
facilities in plan aim to conduct large-scale cosmological
spectroscopic surveys for the $z\sim2-4$ Universe, including the
Subaru Prime Focus Spectrograph \citep[PFS;][]{Takada:2014}, DESI-II,
Stage-V experiments such as Maunakea Spectroscopic Explorer
\citep[MSE;][]{Marshall:2019}, MegaMapper \citep{Schlegel:2019}, and
SpecTel \citep{Ellis:2019}. In the longer wavelength regime, the 50-m
single-dish submillimeter telescope AtLAST plans to perform the first
cosmological spectroscopic surveys for over 100,000 galaxies in
(sub)millimeter in degree square scales \cite{van_Kampen:2024}.

The galaxy populations that these future facilities will target
include Lyman Alpha Emitters (LAEs), Lyman Break Galaxies (LBGs), and
dusty star-forming galaxies (DSFGs), and they are shown to probe halo
masses that differ by up to two orders of magnitude, from
$\sim$10$^{11}$\,M$_\odot$ by the LAEs \citep{Khostovan:2019} to
$\sim$10$^{13}$\,M$_\odot$ by the DSFGs
\citep{Lim:2020,Stach:2021}. The combined measurements of the various
galaxy populations, therefore, allow us to suppress the cosmic
variance via the multi-tracer technique
\cite{McDonald:2009,Ebina:2024}.  Many forecasts have been made for
such tracers to test the precision of upcoming and future cosmological
surveys including the PFS \cite{Takada:2014,Okumura:2022}, the {\rm
  Euclid} space telescope \cite{euclid2022}, MegaMapper, MSE and
MUltiplexed Survey Telescope (MUST)
\cite{d_Assignies_D_2023}. However, the cosmological forecasts were
based on the Fisher matrix formalism, and thus the expected
constraints would be too optimistic. It is necessary to utilize mock
catalogs to provide more realistic forecasts.

In this paper, using dark matter $N$-body simulations run assuming a
spatially-flat $\Lambda$CDM model \cite{Planck-Collaboration:2016}, we
aim to provide a first forecast on the cosmological constraints via
clustering measurements that will be enabled by future spectroscopic
surveys. As a first step, we focus on redshifts of 2 and 3, a natural
extension of current state-of-the-art measurements
\cite{DESI_iv:2025}.  For each snap shot, we construct two dark-matter
halo samples with higher and lower number densities, which roughly
correspond to the PFS ELG survey \citep{Takada:2014} and the AtLAST
\citep{van_Kampen:2024}, respectively. We measure the correlation
functions of dark-matter halos from the two snapshots. We then discuss
how much cosmological information can be extracted from baryon
acoustic oscillations (BAO) and redshift-space distortions (RSD) as
well as the shape of the underlying matter power spectrum.  We
investigate how strongly the cosmological parameters can be
constrained, including the spatial curvature ($\Ok$) and Hubble
constant ($H_0$) from observations at such a high-redshift universe.


{\it $N$-body Simulations and halo catalogs.} We use $N$-body
simulations run as part of an extension of the \darkquest project
\citep{Nishimichi:2019}. We employ $3000^3$ dark matter particles of
mass $m_p=2.584\times 10^{10} h^{-1}M_\odot$ in a cubic box on the
side $2h^{-1}{\rm Gpc}$.  We have the data set from three independent
realizations and we specifically analyze clustering of dark matter
halos of the snapshots at $z=2$ and $z=3$.  Halos are identified using
the \rockstar algorithm \citep{Behroozi:2013}.  Their velocities and
positions are determined by the average of the member particles within
the innermost 10\% of the subhalo radius. The halo mass, $\Mh$, is
defined by a sphere with radius within which the enclosed average
density is 200 times the mean matter density.

The statistical properties of halos that host galaxies at high
redshifts, such as dusty galaxies and emission line galaxies, have not
been accurately determined yet by observations. In this paper, thus,
we do not consider any specific galaxy population but use all subhalos
with some mass range. In order to see the halo mass dependence of the
cosmological results, we adopt two mass ranges as shown in Table
\ref{tab:halos}, referred to as the low- and high-mass samples,
roughly corresponding to the mass ranges of emission line galaxies
($\sim 10^{-3}[\hmpcci]$) \cite{Takada:2014} and dusty galaxies ($\sim
10^{-4}[\hmpcci]$) \citep{van_Kampen:2024}, respectively.


\begin{table}
\caption{Properties of mock halo catalogs. The second column shows the
  mass range of the halos, $\bMh$ is the average mass and
  $\bar{n}_{\rm g}$ is the number density.}
\begin{center}
\begin{tabular}{ccccccc}
\hline\hline
\multirow{2}{*}{$z$} & & \multirow{2}{*}{$\log{\Mh/(\hiMsun)}$} && \multirow{2}{*}{$\log{\bMh/(\hiMsun)}$} && $10^{4}\bar{n}_{\rm g}$  \\ 
   &&  &&&& ($ \hmpcci$)   \\ 
\hline
2 & &$11.5-12.0$ && 11.7 && 32.2 \\
 && $12.5-13.5$ && 12.8 && 4.05 \\
\hline
3 && $11.5-12.5$ && 11.9 && 26.8 \\
 && $12.5-13.5$ && 12.7 && 1.02 \\
\hline\hline
\end{tabular}
\end{center}
\label{tab:halos}
\end{table}


{\it Measurements of redshift-space correlation functions.}  From the
simulations, we measure the redshift-space correlation functions of
halos, $\xi(r,\mu_{\bfr})$, where $r=|\bfr|$ with $\bfr$ being the
separation vector between two points and $\mu_{\bfr}$ is the direction
cosine between the line of sight and $\bfr$.  We then determine the
multipole moments of the correlation functions \citep{Hamilton:1992},
\be
\xi_\ell (r) = (2\ell+1)\int^1_0 \xi(r,\mu_{\bfr}){\cal L}_\ell(\mu_{\bfr})d\mu_{\bfr}, \label{eq:xil}
\ee
where ${\cal L}_\ell(\mu_{\bfr})$ is the $\ell$-th order Legendre
polynomials.

We need to estimate a covariance matrix for the measured correlation
functions for cosmological analysis. We adopt a bootstrap resampling
method \citep[e.g.,][]{Norberg:2009},
\begin{align}
C_{ij} &
\equiv C\left[ \xi_\ell(r_i),\xi_{\ell'}(r_j) \right] \nn \\
&=\frac{1}{N_{\rm mock}-1}
\sum^{N_{\rm mock}}_{k=1}
\left[ \xi_{\ell,k}(r_i)-\bar\xi_{\ell}(r_i) \right] 
\nn \\ & \qquad\qquad\qquad\qquad\times
\left[ \xi_{\ell,k}(r_j)-\bar\xi_{\ell}(r_j) \right],
\end{align}
where $\xi_{\ell,k}(r_i)$ is the measurement from the $k$-th bootstrap
realization and $\bar\xi_\ell(r_i)=N_{\rm mock}^{-1}\sum_{k=1}^{N_{\rm
    mock}}\xi_{\ell,k}(r_i)$.  We create fifty realizations from each
simulation. Furthermore, in measuring the redshift-space density
field, we regard each direction along the three axes of simulation
boxes as the line of sight. Thus, we have $N_{\rm mock}=50\times
3\times 3=450$ realizations to construct the covariance matrix.

\begin{figure}[t]
    \centering
    \includegraphics[width = .49\textwidth]{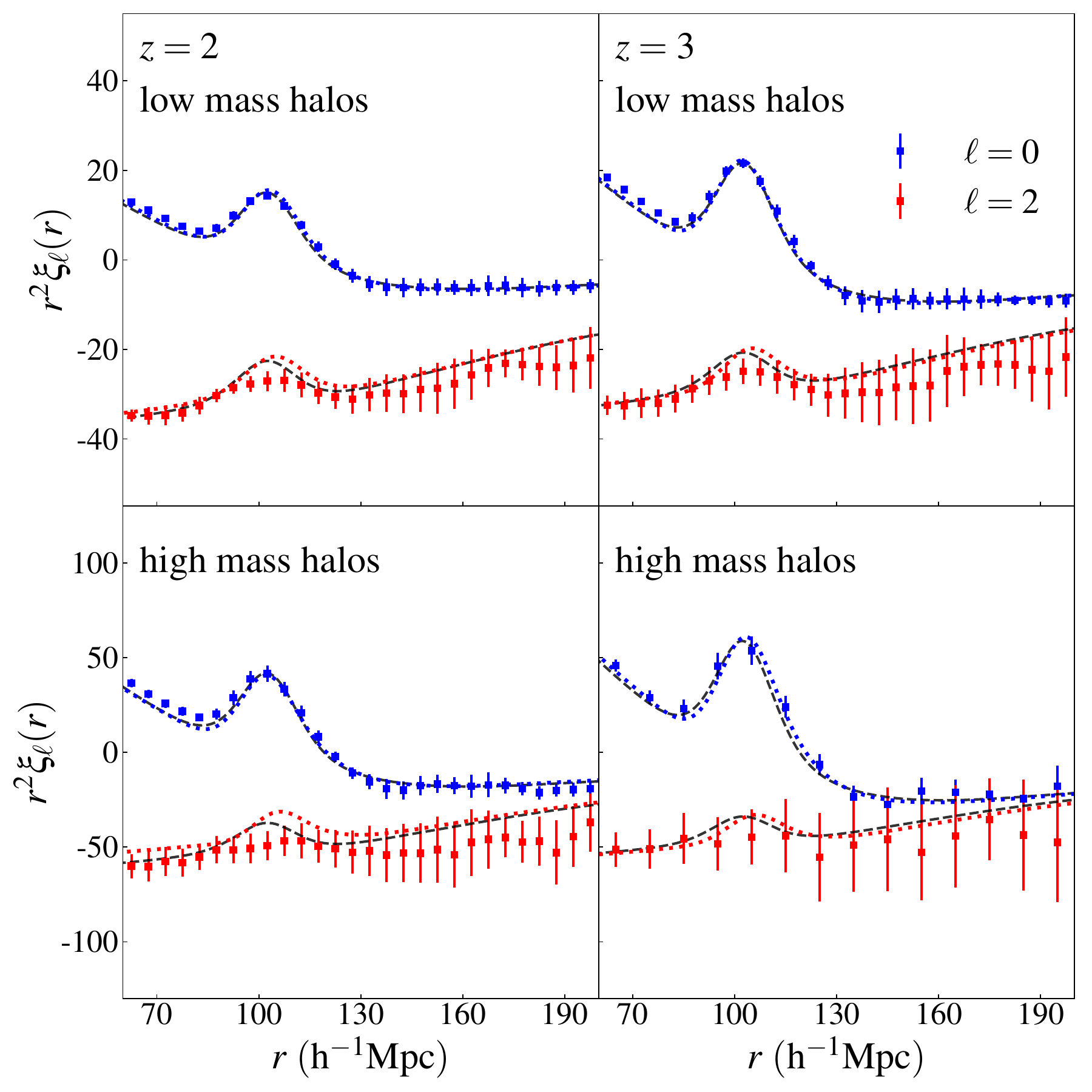}
    \caption{Multipoles of redshift-space correlation functions of
      halos at $z=2$ (left) and $z=3$ (right). The upper and lower
      rows show the results for the low- and high-mass halos,
      respectively. The red and blue points are the measured monopole
      and quadrupole moments from $N$-body simulations.  The dashed
      curves are the model with the input cosmological parameters. The
      dotted curves are the best-fitting model at $60\leq r\leq
      200\himpc$ (see text).  Errors are the square root of the
      diagonal components of the covariance matrices.  }
    \label{fig:tpcf}
\end{figure}

Fig.~\ref{fig:tpcf} shows both the monopole and quadrupole moments of
the redshift-distorted-space correlation functions of redshift 2 and 3
halos for the high-mass and low-mass bins. The error bars for the
multipoles shown are the square root of the diagonal parts of the
covariance matrix.

{\it Theoretical model.} Here we present a theoretical model of the
correlation functions to place cosmological constraints.  Since
theoretical models are derived in Fourier space, we first consider a
model for the redshift-space power spectrum,
$P(\mathbf{k})=P(k,\mu_{\bfk})$, with $k=|\bfk|$ and $\mu_{\bfk}$ the
directional cosine. Given the underlying matter power spectrum in real
space, $P_\mathrm{m}(k)$, cosmological information is extracted from
the dynamical and geometric distortions, which are probed,
respectively, via RSD and Alcock-Paczynski (AP) effects
\citep{Alcock:1979,Kaiser:1987,Seo:2003,Matsubara:2004,Okumura:2008}.

We use the simplest model for RSD, the linear RSD model
with a constant bias $b$ \citep{Kaiser:1987}. It is given by
\begin{equation}
    P(k,\mu_{\bfk}) = (b+f\mu_{\bfk}^2)^2P_\mathrm{m}(k), \label{eq:kaiser}
\end{equation}
where $f(z)$ is the growth rate of the universe sensitive to
modification of the gravity models and approximated by
$f(z)=\Om^{6/11}(z)$ for the $\Lambda$CDM model. The BAO wiggles
encoded in $P_\mathrm{m}(k)$ are damped due to the non-linear structure
formation. 
Density-field reconstruction is widely used for increasing the
precision and accuracy of the BAO detection
\cite{Eisenstein:2007,Paillas:2025}. However, implementing the
technique in full-shape analysis is non-trivial and different
reconstruction methods would yield different constraints
\cite{Zhang:2025}. We instead adopt a conservative approach by
including BAO damping in our model.
Since the damping is less significant at high redshift, we
include it using a simple model (e.g,~\cite{Chuang:2013}),
\begin{equation}
  P_\mathrm{m}(k) = P_\mathrm{nw}(k) + [P_\mathrm{lin}(k)-P_\mathrm{nw}(k)]\exp\left(-\frac{k^2}{2k_*^2}\right),
\end{equation}
where $P_\mathrm{lin}(k)$ and $P_\mathrm{nw}(k)$ are the linear matter
power spectra with and without BAO wiggles, respectively
\cite{Eisenstein:1998}. The parameter $k_*$ controls the effect of the
damping. Although it can be computed analytically \cite{Vlah:2016}, we
simply treat it as a free parameter and marginalize it over to obtain
cosmological constraints.

Next, we consider the geometric distortion, the AP effect, induced by
the apparent mismatch between the reference and true cosmology. This
effect is modeled as
\be
P^{\obs}(k,\mu_{\bfk})=\frac{H}{\Hf}\left(\frac{\DAf}{\DA}\right)^2P(q,\nu_{\bfq}),
\ee
where 
\be
q(k,\mu_{\bfk})=\alpha(\mu_{\bfk})k, \qquad 
\nu_{\bfq}(k,\mu_{\bfk})=\frac{1}{\alpha(\mu_{\bfk})}\frac{H}{\Hf}\mu_{\bfk},
\ee
with $\alpha(\mu_{\bfk})$ being
\be
\alpha(\mu_{\bfk})=\sqrt{\left(\frac{\DAf}{\DA}\right)^2 + \left[ \left(\frac{H}{\Hf}\right)^2- \left(\frac{\DAf}{\DA}\right)^2 \right] \mu_{\bfk}^2 }~.
\ee
Here $H(z)$ and $\DA(z)$ are the expansion rate and angular diameter
distance, respectively. Quantities with the subscript $\rm{fid}$ are
computed using fiducial cosmological parameters (see
Introduction). Detecting anisotropies of BAO enables us to separate
the AP effect from the dynamical distortion effect.

Given $P^\obs(k,\mu_{\bfk})$, the multipole moments are obtained as
\be
P^\obs_\ell (k) = (2\ell+1)\int^1_0 P^\obs(k,\mu_{\bfk}){\cal L}_\ell(\mu_{\bfk})d\mu_{\bfk} \, , 
\label{eq:pkl}
\ee
with $\ell=0,2,4$ containing cosmological information
\cite{Cole:1994,Okumura:2011}.  These multipole power spectra can be
converted into multipole correlation functions by
\begin{equation}
    \xi_\ell(r) = i^\ell\int\frac{dkk^2}{2\pi^2}P^\obs_\ell(k)j_\ell(kr),
\end{equation}
where $j_\ell$ is the $\ell$-th order spherical Bessel function.

In summary, given the matter power spectrum, $P_\mathrm{m}(k)$, the
measured redshift-space correlation function is characterized by a set
of five parameters, $\theta = (b(z),f(z),H(z),\DA(z),k_*)$.  The
prediction with the input cosmological model for our simulations is
shown by the black dashed curve in Fig.~\ref{fig:tpcf}.  Later in the
paper, we perform direct constraints on cosmological parameters
allowing the shape of the underlying power spectrum to vary, referred
to as a full-shape analysis. For this case, the choice of the
parameter space is arbitrary, and the above dynamical and geometric
parameters are computed using the chosen parameters.


{\it Setup for cosmological analysis.} In this paper, we investigate
how well one can constrain cosmological models at redshifts $z>2$ in
two ways: (i) constrain the growth rate and expansion rate using
dynamical and geometric distortions, (ii) constrain cosmological
parameters directly by utilizing the cosmological dependence of the
underlying power spectrum shape.

We perform the likelihood analysis using the measured correlation
functions of the halos with the theoretical models described above.
The $\chi^2$ statistic is given by
\be
\chi^2(\theta) = 
\sum^{N_{\rm bin}}_{i=1} 
\sum^{N_{\rm bin}}_{j=1} 
\Delta_i C_{ij}^{-1}\Delta_j,
\ee
where $\Delta_i\equiv\Delta(r_i;\theta)=\xi_\ell^{\rm
  sim}(r_i)-\xi_\ell^{\rm th}(r_i;\theta)$ is the difference between
the measured and predicted correlation functions, with $\theta$ being
a parameter set to be constrained. The analysis is performed on the
adopted scales $r_{\rm min}\leq r_i\leq r_{\rm max}$ with the number
of bins for all multipole being $N_{\rm bin}$.  The degree of freedom
is $N_{\dof}=N_{\rm bin}-N_p$ with $N_p$ the number of free
parameters.
%
\begin{figure}
    \centering
    \includegraphics[width=.475\textwidth]{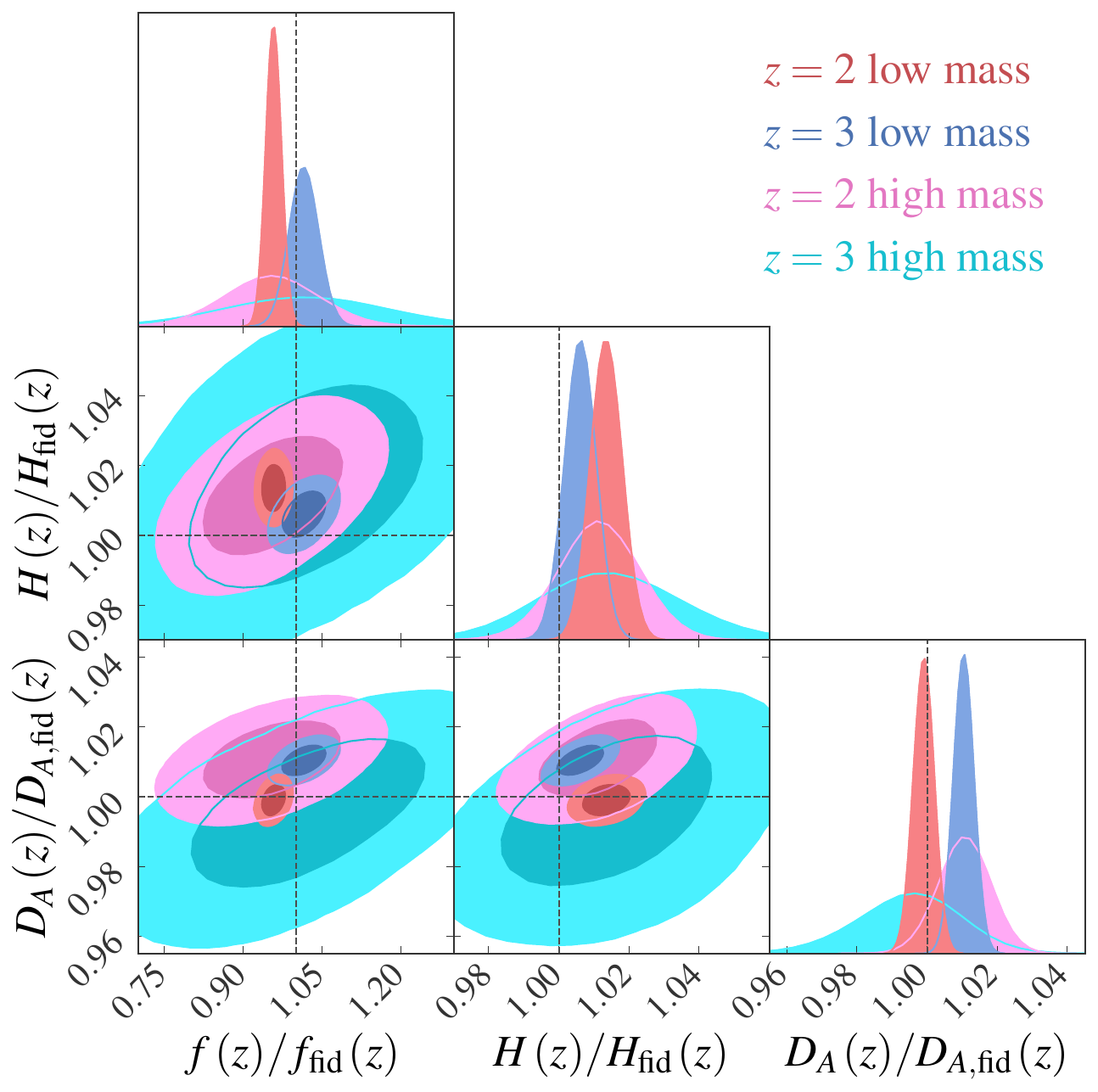}
    \caption{Constraints on dynamical and geometric parameters,
      $(f(z),H(z),\DA(z))$, obtained by the correlation functions of
      low-/high-mass halos at $60\leq r \leq 200\himpc$. Nuisance
      parameters, $b$ and $k_*$, are marginalized over. The contours
      show the 68\% and 95\% C.L. from inward. }
    \label{fig:template}
\end{figure}
The correlation functions on small scales are affected by various
non-linear effects, namely non-linear evolution, RSD, and biasing. On
the other hand, on large scales the covariance matrix is biased due to
our resampling method. Thus, we use a conservative range for the
likelihood analysis and set $(r_{\rm rmin}, r_{\rm max}) =
(60,200)[\himpc\,]$.  For this range, the non-linearity of the matter
power spectrum is completely negligible.  We therefore compute
$P_\mathrm{m}(k)$ in linear theory using the public code
\texttt{CLASS} \citep{Blas:2011}.  Since the nonlinear RSD also does
not significantly affect our result, we can safely use the linear RSD
model (Eq.~\ref{eq:kaiser}). Although the nonlinear bias effect is
severer as we consider relatively massive halos, we still adopt the
linear bias for simplicity.
We set the bin size of the correlation function to $\Delta r =
5\himpc$ for both monopole and quadrupole, thus $N_{\rm bin}=56$, for
all the halo samples except the high-mass halo sample at $z=3$
($\Delta r = 10\himpc$ and $N_{\rm bin}=28$) that is so sparse that
the correlation function becomes noisy.
To perform a maximum likelihood analysis we
use the Markov Chain Monte Carlo sampler \texttt{emcee}
\citep{EMCEE:2013}, under flat priors otherwise stated.

\begin{figure}[t]
    \centering
    \includegraphics[width=.4\textwidth]{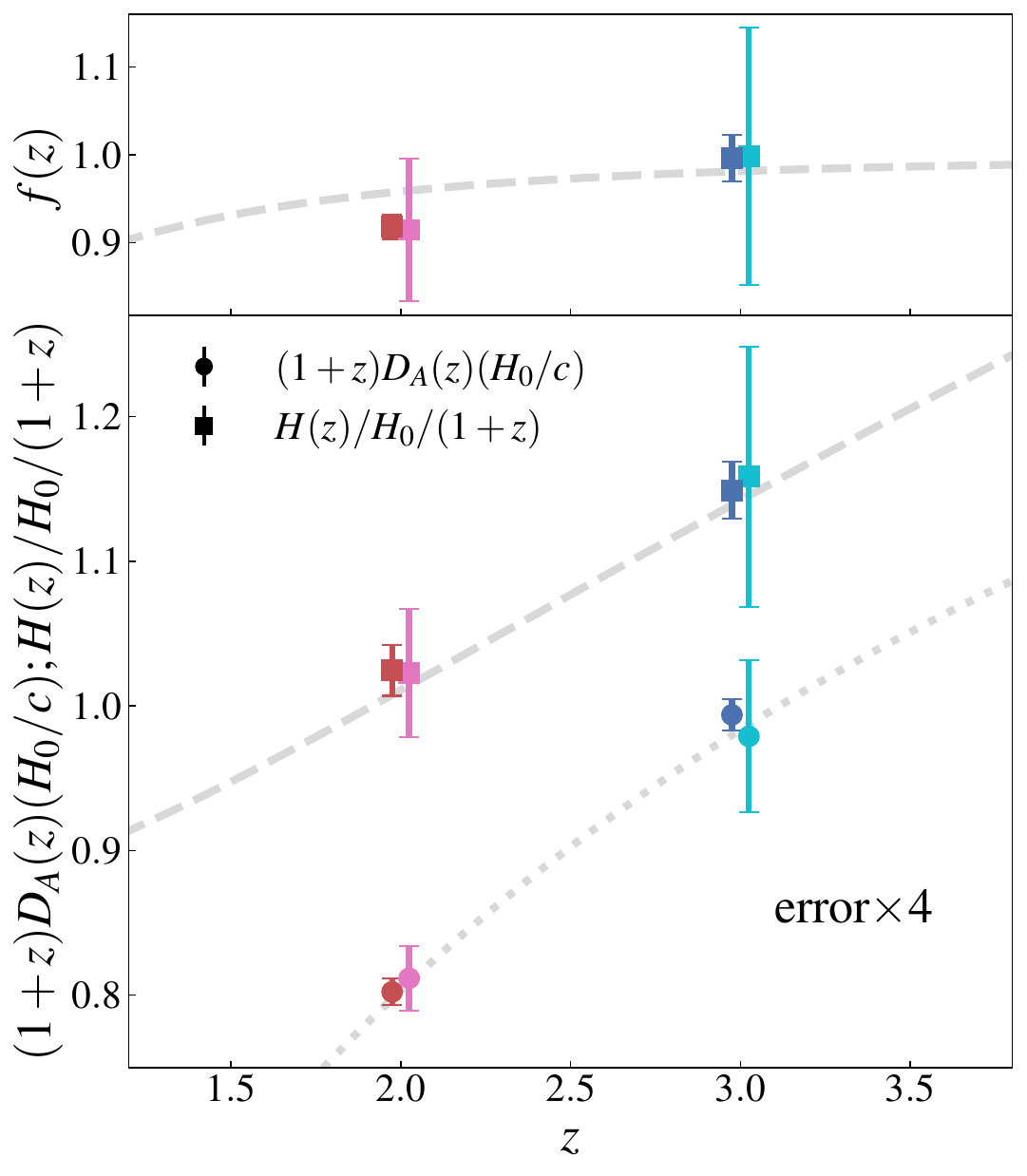}
    \caption{One-dimensional marginalized errors on $f(z)$ (upper
      panel) and $H(z)$ and $\DA(z)$ (lower panel) as functions of
      redshift. Gray curves are the true values of parameters at
      redshift 2 and 3. Errors of 68\% confidence level are shown, but
      those on $H$ and $\DA$ are multiplied by 4 for illustration. For
      clarity, the results from the low- and high-mass halos are
      slightly offset horizontally toward left and right,
      respectively.}
    \label{fig:redshift}
\end{figure}

Cosmological constraints obtained below correspond to those expected
from a galaxy survey of the volume of our simulations,
$V=(2\higpc)^3$, smaller than typical upcoming surveys. However, since
data in actual surveys are split into redshift bins for cosmological
analysis, our constraints will roughly correspond to those from one
redshift bin.

{\it Dynamical and geometric constraints.} For the analysis of
dynamical and geometric distortions, we have five parameters in total,
$\theta=(b,f,H,\DA,k_*)$, among which $b$ and $k_*$ are nuisance
parameters that we want to marginalize over.  The matter power
spectrum is computed assuming the fiducial cosmology.

\begin{table*}
\caption{Dynamical and geometric parameter constraints. }
\begin{center}
\begin{tabular}{ccccccccccccc}
\hline\hline
& &\multicolumn{5}{c}{$z=2$}  && \multicolumn{5}{c}{$z=3$}    \\  
 \cline{3-7}
 \cline{9-13}
$\theta$ && fiducial && low mass && high mass & & fiducial && low mass& & high mass  \\ 
\hline
$f$&& 0.958 & &$0.918^{+0.014}_{-0.013}$ && $0.914^{+0.082}_{-0.082}$& & 0.981 && $0.996^{+0.027}_{-0.027}$ && $0.998^{+0.147}_{-0.147}$ \\ 
$\DA$&& 1193.3 & & $1192.1^{+3.4}_{-3.4}$&& $1206.0^{+8.4}_{-8.6}$ & & 1096.0 && $1107.4^{+3.0}_{-3.0}$ && $1090.9^{+14.6}_{-16.1}$ \\ 
$H$&& 204.09 && $206.85^{+0.89}_{-0.86}$ && $206.44^{+2.24}_{-2.23}$ & &307.43&& $309.31^{+1.32}_{-1.32}$ && $311.84^{+6.06}_{-5.76}$ \\ 
$b$& &&& $1.942^{+0.015}_{-0.015}$ & &$3.365^{+0.099}_{-0.103}$ & &&&  $3.238^{+0.025}_{-0.025}$ && $5.514^{+0.176}_{-0.178}$ \\ 
$\chi_{\rm min}^2/N_{\dof}$ &&&& $206.063/51$ && $48.069/51$ &&&& $129.219/51$ && $14.976/23$ \\
\hline\hline
\end{tabular}
\end{center}
\label{tab:params1}
\end{table*}

Fig.~\ref{fig:template} shows two-dimensional error contours on the
geometric and dynamical parameters for the four samples, normalized by
their fiducial values. Fig.~\ref{fig:redshift} shows the
one-dimensional marginalized errors as a function of redshift.  The
constrained parameters are summarized in Table \ref{tab:params1}.  The
preliminary results of this analysis have been presented in
\citep{van_Kampen:2024}. The best-fit model obtained here is shown by
the dotted curves in Fig.~\ref{fig:tpcf}.

Since the simulation box size is fixed, the strength of the
constraints depends on the number density of halo samples.  The
best-fit models are consistent with the input (true) models of the
simulations within $2\sigma$ level except for the low-mass sample at
$z=2$. This offset is caused by the fact that our simple linear model
is not very accurate for the precision measurement.  Furthermore, this
model provides a poor fit for this sample with the reduced $\chi^2$
value of $\chi_{\rm min}^2/N_{\dof} \approx 4$. However, such
systematic effects can be easily controlled once a more sophisticated
model is adopted to take account of nonlinear corrections (e.g.,
\cite{Nishimichi:2020,Chen:2021}). The precision of the growth rate is
at most $15\%$ for low number density and high mass halos. The
precision is significantly improved to $\sim 2-3\%$ as the number
density increases to a few times $10^{-3}\hmpcci$.  We see a similar
trend for $H(z)$ and $\DA(z)$, whose constraints improve from $1.9\%$
to $0.4\%$ and $1.3\%$ to $0.3\%$, respectively.  The Lyman-$\alpha$
survey result in the DESI Collaboration \cite{DESI_iv:2025} gives
$2\%$ and $2.4\%$ precision in $H(z)$ and $\DA(z)$, respectively, with
number density $n_{\rm g} = 3\times 10^{-5}\hmpcci$ at effective
redshift 2.33. Our results at $z=2$ are, therefore, confirmed by
observations.

{\it Full-shape constraints.} Here we utilize the cosmological
information encoded in the underlying matter power spectrum and
directly constrain cosmological parameters. We consider a nonflat
$\Lambda$CDM model with six parameters in total,
$\theta=(b,~\Om,~\Ok,~H_0,~\As,~k_*)$, where $\As$ is the amplitude of
the primordial curvature perturbations. We fix the baryon density,
$\ob\equiv\Ob h^2$, to the fiducial value because it is tightly
constrained by CMB and big-bang nucleosynthesis
\cite{Aver:2013,Cooke:2014,Planck-Collaboration:2016}. We further fix
the dark energy equation-of-state to a cosmological constant,
$w_0=-1$, as the high-$z$ clustering is not sensitive to $w$. On the
other hand, the high-redshift universe is a useful laboratory to test
the flatness of the universe, $\Ok$ \citep{Takada:2015}.

Since there is a strong degeneracy between $b$ and $\As$ in linear
theory, they cannot be tightly constrained \cite{Okumura:2008}.  For
sampling stability and efficiency of the MCMC, we adopt a Gaussian
prior on $\As$ and marginalize it over together with the other
nuisance parameters, $b$ and $k_*$.  We confirmed that this does not
affect constraints on the other parameters, since the degeneracies
between $\As$ and other parameters are relatively small.\footnote{
Using the shape information of the power spectrum on nonlinear scales
and the bispectrum enables one to simultaneously determine $b$ and the
perturbation amplitude, and thus to investigate the $\sigma_8$ (or
$S_8$) tension \cite{Philcox:2022}. Such an investigation is left to
our future work.} Here we exclude $z=2$ low-mass halo samples from our
discussion, as our model of the correlation function is too simple to
provide accurate cosmological constraints via the full-shape analysis
from them.

\begin{figure}[b]
    \centering
    \includegraphics[width=\linewidth]{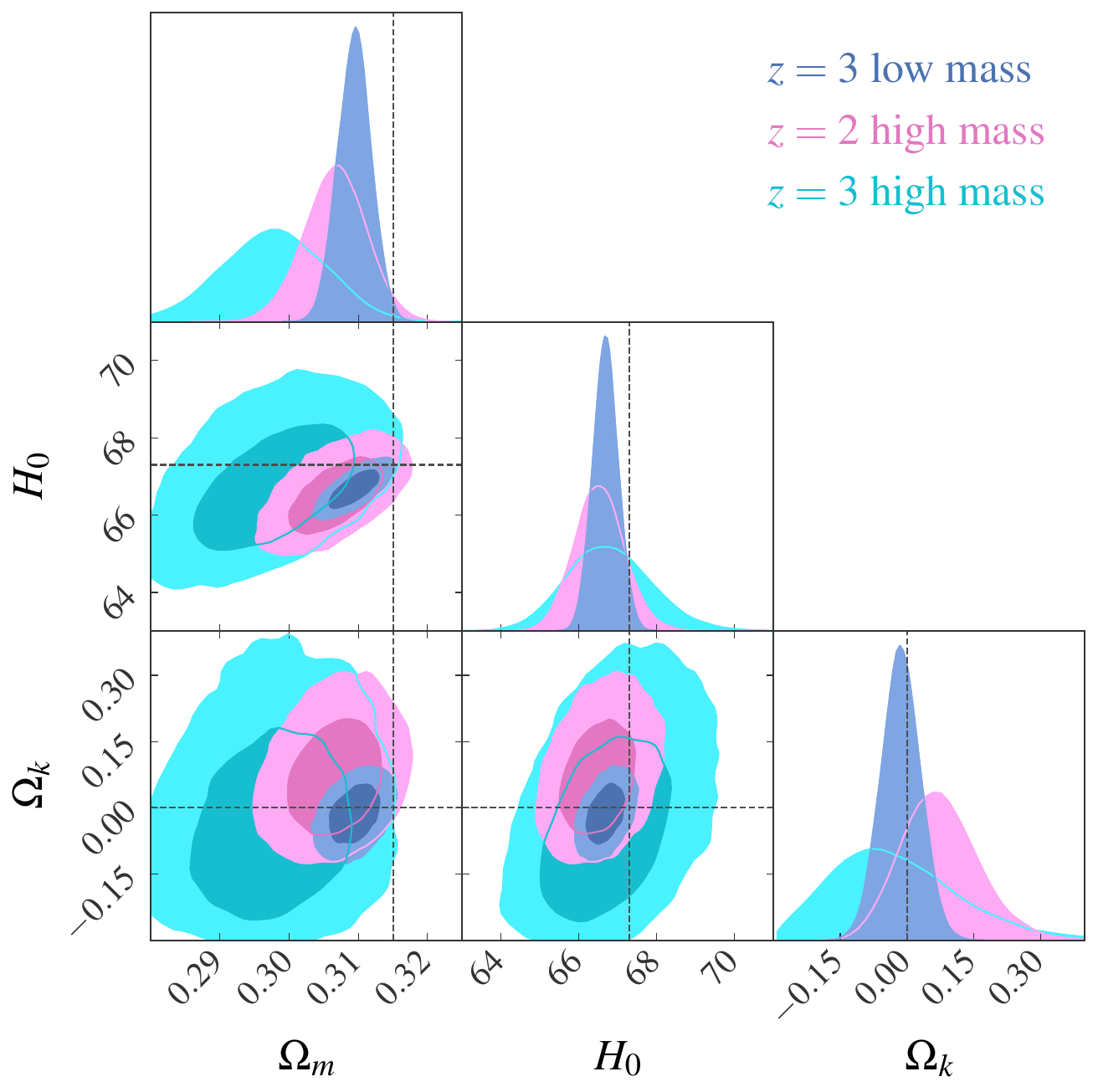}
    \caption{Constraints on cosmological parameters $(\Om,H_0,\Ok)$
      from the full-shape information of correlation functions of
      high-mass halo samples at $z=2$, as well as low-/high-mass halo
      samples at $z=3$. Nuisance parameters, $b$ and $k_*$, as well as
      the amplitude $\As$ are marginalized over. The dotted lines
      indicate the fiducial values.  }
    \label{fig:cosmological_params}
\end{figure}

Fig.~\ref{fig:cosmological_params} shows the two-dimensional
constraints on pairs of cosmological parameters. Table
\ref{tab:params2} summarizes the one-dimensional marginalized
constraints. The best-fitting value of $b$ obtained here is fully
consistent with that from dynamical and geometric constraints.  We
obtain $0.7-2.5\%$ constraints on $\Om$ and $0.5-1.7\%$ on $H_0$. The
reported discrepancy between the values of $H_0$ measured using early
and late universe probes is about $5-10\%$ \cite{Verde:2019}. Our
result thus demonstrates that achieving the number density of
$\bar{n}_{\rm g}\simeq 10^{-4} (\hmpcci)$ over the large volume at
$z\geq 2$ provides a powerful test of the Hubble tension, independent
of the local universe ($z\simeq 0$) and early universe ($z\simeq
1100)$ measurements.

Note that there are certain offsets for the constraint on $\Om$ from
the $z=2$ sample, though $\simeq 2\sigma$, while the results shown in
the previous subsection are consistent with the input within the
$1\sigma$ level for these samples.  These shifts are driven by the
inaccuracy of our linear theory model with the simple BAO damping on
small scales, where cosmological information can still be extracted.
In fact, Fig.~\ref{fig:tpcf} shows that the best-fit model fails to
capture behavior on smaller scales than the BAO scale, especially at
$z=2$, while it agrees with the measurements up to the BAO scale. This
issue can be easily restored once one employs a more sophisticated
model, and our main conclusion of constraining cosmology from high
redshifts is unchanged.  Small discrepancies of the quadrupoles around
the BAO scales will be further improved by including the
angle-dependent BAO damping.


\begin{table}
\caption{Cosmological parameter constraints.}
\begin{center}
\begin{tabular}{cccccccccc}
\hline\hline
& & & $z=2$  & & & \multicolumn{3}{c}{$z=3$}    \\  
\cline{3-5}
\cline{7-9}
$\theta$ & fiducial & &high mass & & & low mass & & high mass  \\ 
\hline
$\Om$ & 0.3150 &&  $0.3067^{+0.0043}_{-0.0045}$ & & & $0.3095^{+0.0023}_{-0.0024}$&&$0.2977^{+0.0074}_{-0.0076}$ \\
$H_0$ & 67.30 &&  $66.52^{+0.63}_{-0.62}$ & & & $66.68^{+0.32}_{-0.32}$&&$66.74^{+1.12}_{-1.02}$ \\
$\Ok$ & 0 &&  $0.070^{+0.088}_{-0.080}$ & & & $-0.014^{+0.044}_{-0.043}$&&$-0.039^{+0.158}_{-0.125}$ \\
$10^9\As $ & 2.12&  & $2.12^{+0.11}_{-0.11}$ & & & $2.14^{+0.11}_{-0.11}$ &&$2.12^{+0.10}_{-0.10}$ \\
$b$ &&  & $3.373^{+0.227}_{-0.218}$ & & & $3.106^{+0.087}_{-0.086}$&&$5.170^{+0.517}_{-0.452}$ \\
$\chi_{\rm min}^2/N_{\dof}$&&  & $54.557/50$  & & & $123.761/50$ &&$11.985/22$ \\
\hline\hline
\end{tabular}
\end{center}
\label{tab:params2}
\end{table}

We also obtain interesting constraints on the spatial curvature at
both redshifts.  Using the Fisher matrix formalism,
Ref.~\cite{Takada:2015} showed that one can achieve the precision of
$|\Ok|\lesssim 3\times 10^{-3}$ (68\%~C.L.) by observing all the
galaxies with $10^{-4}(\hmpcci)$ at $0\leq z \leq z_{\rm max}$ where
$z_{\rm max}\simeq 2$, which is the most optimistic and ideal
situation.  With more realistic simulation-based analysis, we show
that the galaxy clustering at one-single redshift of $z=2$ or $z=3$
provides a constraint on $\Ok$, as $|\Ok|\lesssim 0.1$, without
relying on other observations such as the CMB.


{\it Summary and discussion.} In this paper, we have studied the
cosmological constraints expected from upcoming high-redshift galaxy
surveys utilizing high-resolution $N$-body simulations at $z=2$ and
3. To obtain general conclusions, we did not consider specific galaxy
populations or survey geometries but rather split our halo sample into
low-mass and high-mass subsamples in a cubic box of side $2\higpc$.
We chose a conservative fitting range where a simple linear theory
prediction recovered the input cosmological models.  We considered two
analysis methods, a template fitting for dynamical and geometric
constraints and a full-shape analysis.  With the former, we obtained a
$\sim 15\%$ constraint on $f(z)$ with the massive halo sample with
$\bar{n}_{\rm g}\sim 10^{-4}\hmpcci$ but the constraint improved to a
few percent by increasing the number density by an order of
magnitude. On the other hand, we achieved $<2\%$ constraints on $H(z)$
and $\DA(z)$ even with the low-number density halo sample.  Our
full-shape analysis demonstrated that we can simultaneously
investigate the Hubble tension and the flatness of the
universe. Including the nonlinearity of the matter spectrum further
enables us to address the $\sigma_8$ (or $S_8$) tension without
combining the CMB priors.

Our analyses provide a timely forecast for the upcoming spectroscopic
surveys. For example, the expected number density for the emission
line galaxy (ELG) sample in PFS at redshift 2 is $2.7\times
10^{-4}\hmpcci$\citep{Takada:2014}. With already good precision in
DESI's results including galaxies spanning on a larger range of
redshifts with a much lower number density of our samples, we can
expect to see percent or sub-percent level constraints on cosmological
parameters in PFS or future DESI releases that have higher number
density. Our forecasts on massive halos are also helpful in guiding
the design of the AtLAST spectroscopic surveys, which will
predominantly target dusty star-forming galaxies at $z\gtrsim2$. The
expected number density obtained from AtLAST is comparable to those of
both of our high-mass bins, which, based on our results, will allow
for percent levels of constraints on cosmological parameters.

This paper did not consider any specific galaxy sample or survey
geometry.  For a more realistic cosmological forecast based on
simulations, we need to generate light-cone output using multiple
snapshots of $N$-body simulations
(e.g.,~\cite{Howlett:2015a,Ishikawa:2024}). Although the halo
occupation distribution (HOD) of ELGs has been studied recently
(e.g.,~\cite{Avila:2020,Okumura:2021}), that of dusty galaxies still
contains several uncertainties. Thus, a cosmological forecast for
dusty galaxies at high redshifts requires more effort. Such detailed
investigations and analyses will be performed in our future work.

{\it Acknowledgments.} TO acknowledges support of the Taiwan National
Science and Technology Council under Grants No. NSTC
112-2112-M-001-034-, NSTC 113-2112-M-001-011- and NSTC
114-2112-M-001-004-, and the Academia Sinica Investigator Project
Grant No. AS-IV-114-M03 for the period of 2025–2029. SYP and CCC
acknowledge the support of the Taiwan National Science and Technology
Council (111-2112M-001-045-MY3), as well as the Academia Sinica
through the Career Development Award (AS-CDA-112-M02). TN acknowledges
support by JSPS KAKENHI Grant Numbers JP20H05861, JP21H01081,
JP22K03634, JP24H00215 and JP24H00221. KA is supported by Fostering
Joint International Research (B) under Contract No. 21KK0050 and the
Japan Society for the Promotion of Science (JSPS) KAKENHI Grant
No. JP24K17056. We acknowledge
\href{https://hpc.tiara.sinica.edu.tw/}{`the HPC facility at ASIAA'}
where part of the numerical analyses were done. This work used
high-performance computing facilities operated by the Center for
Informatics and Computation in Astronomy (CICA) at National Tsing Hua
University. This equipment was funded by the Ministry of Education of
Taiwan, the Ministry of Science and Technology of Taiwan, and National
Tsing Hua University. We are grateful to the maintenance and
administrative staff of our institutions, whose efforts in supporting
our day-to-day work environment make our scientific discoveries
possible.


\bibliography{ms_prd.bbl}

\end{document}